# Some Properties of Brzozowski Derivatives of Regular Expressions

N.Murugesan [#1], O.V.Shanmuga Sundaram [*2]

[#1] Assistant Professor, Dept of Mathematics, Government Arts College (Autonomous),
Coimbatore – 641018, Tamil Nadu, India,
[*2] Assistant Professor, Dept of Mathematics, Sri Shakthi Institute of Engineering & Technology,
Coimbatore – 641062, Tamil Nadu, India.

***Abstract*** **:- Brzozowski's derivatives of a regular expression are developed for constructing deterministic automata from the given regular expression in the algebraic way. In this paper, some lemmas of the regular expressions are discussed and the regular languages of the derivatives are illustrated. Also the generalizations of the Brzozowski's derivatives are proved as theorems with help of properties and known results.**

## I. Introduction

Regular expressions are declarative way of defining regular languages recognized by a DFA or a NFA. They are equivalent to one another in the sense that, for a given regular expression, it can be constructed a finite state automata recognizing the same language described by the regular expression, and vice–versa.

All over the years, various attempts have been made to accomplish this task. In the year 1960, R.McNaughton and H.Yamada [6] provided an algorithm to construct a non – deterministic finite automaton from a regular expression.

G.Berry and R.Sethi [1] discussed the theoretical background for the R.McNaughton and H.Yamada algorithm. V.M.Glushkov [4] has also given a similar algorithm in the year 1961. An elegant construction of deterministic finite automata based on the derivatives of regular expressions was proposed by J.A Brzozowski [2] in the year 1964. J.E.Hopcroft and J.D.Ullman [5] discussed the construction of $\varepsilon$ - NFA from the given regular expression. J.M.Champarnaud and others [3] described a variant of the step by step construction which associates standard and trim automata to regular languages.

In this paper, we discuss some basic set theoretic properties involved in Brzozowski way of constructions of automata have been discussed.

## II. Regular expressions

Let $\sum$ be an alphabet of symbols. A word over an alphabet $\sum$ is a finite sequence of symbols from that alphabet $\sum$. The set of all words over $\sum$ is denoted by $\sum^*$. The empty word is denoted by $\varepsilon$. A regular expression is defined inductively as

(i) $\varepsilon$ is a regular expression.

(ii) For any $a \in \sum$, the symbol 'a' is a regular expression.

(iii) If E and F are regular expressions, then $E+F, EF, (E)^*$ are all regular expressions.

The regular expressions $E+F$, $EF$, $(E)^*$ are called respectively union, concatenation, Kleene closure of the corresponding regular expressions. The language of a regular expression E is denoted as $L(E)$, and defined the same for various regular expressions as follows.

(i) $L(\varepsilon)=\{\varepsilon\}$ (ii) $L(a)=\{a\}$

(iii) $L(E+F)=L(E)\cup L(F)$

(iv) $L(EF)=L(E)L(F)$ (v) $L(E^*)=(L(E))^*$

The empty set $\varnothing$ is also considered as a language of regular expression denoted by the symbol $\varnothing$ itself. It is assumed that

$$E+\varnothing=\varnothing+E=E; \quad E\varnothing=\varnothing E=\varnothing; \quad \varepsilon E=E\varepsilon=E$$

The properties of the regular languages are discussed in [9].

The following lemma gives some algebraic type identities with respect to regular expressions.

*2.1 Lemma*

Let E and F are any two regular expressions. Then,

$(i)$ $E+F=F+E$

$(ii)$ $EF=FE$ *only when*

$(a)$ $E=F$ *or* $(b)$ *one of* $E,F$ *is* $\varepsilon$ *or* $\varnothing$.

$(iii)$ $(E+F)+G=E+(F+G)$

$(iv)$ $(E^*)^*=E^*$

$(v)$ $E^*(F+G)=E^*F+E^*G$

Not all algebraic type identities are hold in the case of regular expressions.

2.2 *Lemma*:

For any two regular expressions E and F, then,

$$(i)\quad (E+F)^* \neq E^* + F^*$$

$$(ii)\quad (EF)^* \neq E^*F^* \qquad (iii)\quad EF \neq FE$$

Provided that E and F are not equal to $\varepsilon$ or $\varnothing$ .

2.3 *Lemma*:

$$(i)\quad (\varepsilon+a)^* = a^* \qquad (vi)\quad \varepsilon + aa^* = a^*$$

$$(ii)\quad a^*(\varepsilon+a) = a^* \quad (vii)\quad (a+b)^* = (a^*b^*)^*$$

$$(iii)\quad (\varepsilon+a)+a^* = a^* \quad (viii)\quad \varnothing a = a\varnothing = \varnothing$$

$$(iv)\quad b+a^*b = a^*b \qquad (ix)\quad \varnothing + a = a + \varnothing = a$$

$$(v)\quad b+ba^* = ba^* \qquad (x)\quad \varepsilon + a^* = a^*$$

$$(xi)\quad a(b+c) = ab+ac \qquad (xiv)\quad \varepsilon a = a\varepsilon = a$$

$$(xii)\quad (a+b)c = ac+bc \qquad (xv)\quad \varepsilon^* = \varepsilon$$

$$(xiii)\quad (a+b)^* = (a^*+b^*)^* = (a^*b^*)^*$$

Some of the proofs of the equivalent regular expressions given in the above lemmas are proved in [7].

## III Derivatives of Regular Expressions

3.1 *Definition*

Given a language L and a symbol 'a', the derivative of L with respect to a symbol a is defined as

$$D_a(L) = \{b \mid ab \in L\}.$$

The derivatives of regular expressions with respect to a symbol are defined as follows:

1 $D_a(\varnothing) = D_a(\varepsilon) = \varnothing \qquad D_\varepsilon(a) = a$

2 $D_a(b) = \begin{cases} \varepsilon & \text{if } b = a \\ \varnothing & \text{Otherwise} \end{cases}$ .

3 $D_a(E+F) = D_a(E) + D_a(F)$

4 $D_a(EF) = \begin{cases} D_a(E)F + D_a(F) & \text{if } \varepsilon \in L(E) \\ D_a(E)F & \text{otherwise} \end{cases}$

5 $D_a(E^*) = D_a(E)E^*$

6 $D_\varepsilon(E) = E$ 7 $D_{wa}(E) = D_a(D_w(E))$

The operator D is treated as a prefix operator with high precedence than "+", "." and "*". The derivatives involving the operators intersection, and complement are defined by

$$D_a(E \cap F) = D_aE \cap D_aF$$

and $D_a(E-F) = D_aE - D_aF$ .

It can be verified that

$$D_a(ab^* \cap a) = D_a(ab^*) \cap D_a(a)$$

$$D_a(ab^* - a) = D_a(ab^*) - D_a(a)$$ .

3.2 *Examples*

1 $\quad Let\ E = a(a+b)^*$

$\quad Then\ D_a(E) = (a+b)^*$

$\quad D_b(E) = \varnothing$

2 $\quad Let\ E = ab(a+b)^*$

$\quad Then\ D_a(E) = b(a+b)^*$

$\quad D_b(E) = \varnothing$

3 $\quad Let\ E = (a+b)^* a$

$$\begin{aligned} Then\ D_a(E) &= \left[D_a\left((a+b)^*\right)\right]a + D_a(a) \\ &= \left[\left[D_a((a+b))\right](a+b)^*\right]a + \varepsilon \\ &= \left[D_a(a) + D_a(b)\right](a+b)^* a + \varepsilon \\ &= (\varepsilon + \varnothing)(a+b)^* a + \varepsilon \\ &= \varepsilon(a+b)^* a + \varepsilon \\ &= (a+b)^* a + \varepsilon \end{aligned}$$

$$\begin{aligned} D_b(E) &= \left[D_b\left((a+b)^*\right)\right]a + D_b(a) \\ &= \left[\left[D_b((a+b))\right](a+b)^*\right]a + \varnothing \\ &= \left[D_b(a) + D_b(b)\right](a+b)^* a \\ &= (\varnothing + \varepsilon)(a+b)^* a \\ &= \varepsilon(a+b)^* a \\ &= (a+b)^* a \end{aligned}$$

.

3.3 *Definition*

Let L be a regular language. We define

$$\delta(E) = \begin{cases} \varepsilon & if\ \ \varepsilon \in L \\ \varnothing & if\ \ \varepsilon \notin L \end{cases}$$

It can be easily seen that

$$(i)\quad \delta(a) = \varnothing, \quad for\ any\ \ a \in \Sigma$$

$$(ii)\quad \delta(\varepsilon) = \varepsilon, \quad and\ \ \delta(\varnothing) = \varnothing$$

$$(iii)\quad \delta(E+F) = \delta(E) + \delta(F)$$

$$(iv)\quad \delta(E^*) = \varepsilon$$

3.4 *Definition*

Let $w = a_1a_2....a_n$ and E be a regular expression. Then,

$$D_{a_1a_2}(E) = D_{a_2}(D_{a_1}(E))$$

$$D_{a_1a_2a_3}(E) = D_{a_3}(D_{a_1a_2}(E))$$

In general, we have

$$D_w(E) = D_{a_1a_2a_3....a_n}(E)$$
$$= D_{a_n}\left(D_{a_1a_2.....a_{n-1}}(E)\right)$$

*3.5 Theorem*

Let E, F are two regular expressions and the word $w = a_1a_2....a_n$ .a string over the Kleene closure of an alphabet $\sum$. Then,

$$D_w(E+F) = D_w(E) + D_w(F)$$

*3.6 Theorem*

Let E, F are two regular expressions and the word $w = a_1a_2....a_n$. Then,

$$D_{a_1}(EF) = \left(D_{a_1}(E)\right)F + \delta(E)D_{a_1}(F)$$
$$D_{a_1a_2}(EF) = \left(D_{a_1a_2}(E)\right)F + \delta(E)D_{a_1}(E)D_{a_1}(F)$$
$$+\delta(E)D_{a_1a_2}(F)$$

In general,

$$D_{a_1a_2....a_n}(EF) = \left(D_{a_1a_2....a_n}(E)\right)F$$
$$+\delta\left(D_{a_1a_2....a_{n-1}}(E)\right)D_{a_n}(F)$$
$$+\delta\left(D_{a_1a_2....a_{n-2}}(E)\right)D_{a_{n-1}a_n}(F)$$
$$+\delta\left(D_{a_1a_2....a_{n-3}}(E)\right)D_{a_{n-2}a_{n-1}a_n}(F)$$
$$+.....+\delta\left(D_{a_1}(E)\right)D_{a_2a_3.....a_n}(F)$$
$$+\delta(P)D_{a_2a_3.....a_n}(F)$$

*3.7 Theorem*

Let the word $w = a_1a_2....a_n$. Then,

$$D_{a_1}(E^*) = \left(D_{a_1}(E)\right)E^*$$
$$D_{a_1a_2}(E^*) = \left(D_{a_1a_2}(E)\right)E^*$$
$$+\delta\left(D_{a_1}(E)\right)D_{a_2}(E^*)$$

$$D_{a_1a_2a_3}(E^*) = \left(D_{a_1a_2a_3}(E)\right)E^*$$
$$+\delta\left(D_{a_1}(E)\right)D_{a_2a_3}(E^*)$$
$$= \left(D_{a_1a_2a_3}(E)\right)E^* + \delta\left(D_{a_1}(E)\right)$$
$$\times\left[D_{a_2a_3}(E)E^* + \delta\left(D_{a_2}(E)\right)D_{a_3}(E^*)\right]$$
$$= \left(D_{a_1a_2a_3}(E)\right)E^* + \delta\left(D_{a_1}(E)\right)D_{a_2a_3}(E)E^*$$
$$+\delta\left(D_{a_1}(E)\right)\delta\left(D_{a_2}(E)\right)D_{a_3}(E^*)$$

Similarly, we can generalize

$$D_{a_1a_2a_3.....a_n}(E^*) = \left(D_{a_1a_2a_3....a_n}(E)\right)E^*$$
$$+\delta\left(D_{a_1}(E)\right)D_{a_2a_3....a_n}(E)E^*$$
$$+....+\delta\left(D_{a_1}(E)\right)\delta\left(D_{a_2}(E)\right)....\delta\left(D_{a_{n-1}}(E)\right)D_{a_n}(E^*)$$

*3.8 Lemma*

$$D_a\left[(aw)^*\right] = \varepsilon + w(aw)^*, \quad where \;\; a \in \Sigma, \; w \in \Sigma^*$$

Proof:

$$L\left[(aw)^*\right] = \{\varepsilon, aw, awaw, awawaw, ....\}$$
$$D_a\left(L\left[(aw)^*\right]\right) = \{\varepsilon, w, waw, wawaw, ....\}$$
$$= \{\varepsilon\} \cup \{w\}\{\varepsilon, aw, awaw, ....\}$$
$$= \{\varepsilon\} \cup w(aw)^*$$

Hence the corresponding regular expression is $\varepsilon + w(aw)^*$.

*3.9 Lemma*

Let E be regular expression, then

$$D_a\left[L(aE)^*\right] = L(E)L(aE)^*, \quad where \;\; a \in \Sigma.$$
$$L(aE)^* = \{\varepsilon, aE, aEaE, ....\}$$
$$D_a\left[L(aE)^*\right] = \{E, EaE, ....\}$$
$$= \{E\}\{\varepsilon, aE, aEaE, ....\}$$
$$= L(E)L(aE)^*$$

*3.10 Theorem*

Let E be any regular expression and a be any symbol over the alphabet $\sum$.

$$L\left(D_a(E)\right) = D_a\left(L(E)\right).$$

Proof:

*Case (i):* Let $E = \varepsilon$, then $L(E) = \{\varepsilon\}$., and $D_a\left(L(E)\right) = \varnothing$.

Also $D_a(\varepsilon) = \varnothing$, and $L\left(D_a(\varepsilon)\right) = \varnothing$.

On the other hand, if $E = \varnothing$, then

$$L\left(D_a(E)\right) = D_a\left(L(E)\right) = \varnothing.$$

Hence the theorem is true when $E = \varepsilon$ and $E = \varnothing$.

*Case (ii):* Let $E = a$.

Then, $D_a(a) = \varepsilon$. Hence $L\left(D_a(E)\right) = \{\varepsilon\}$.

Also $L(E) = \{a\}$., and $D_a\left(L(E)\right) = \{\varepsilon\}$.

If $E = b \neq a$, then $L(D_a(E)) = D_a(L(E)) = \varnothing$.

*Case (iii):* Let $E = F + G$ . We prove first $D_a(E) = D_a(F) + D_a(G)$.

Suppose, if $F = aF'$ and $G = aG'$ ,then $D_a(E) = F' + G'$ .and

$L(D_a(E)) = L(F') \cup L(G')$

Let F and G are two regular expressions begin with a symbol other than 'a'. Then

$D_a(F) = \varnothing = D_a(G)$

Hence $D_a(E) = \varnothing$ and $L(D_a(E)) = \varnothing$ .

In the first case, $L(F) = \{aw \mid w \in \Sigma^*\}$,

and $L(G) = \{au \mid u \in \Sigma^*\}$ .

Hence $L(E) = \{aw, au \mid w, u \in \Sigma^*\}$.

Hence $L(E) = aL(F') \cup aL(G')$.

i.e., $D_a(L(E)) = L(F') \cup L(G')$.

In the second case, $L(F) = \{bw \mid b \neq a, w \in \Sigma^*\}$

.and $L(G) = \{bu \mid b \neq a, u \in \Sigma^*\}$.

$L(E) = \{bw, bu \mid b \neq a, w, u \in \Sigma^*\}$.

Therefore, $D_a(L(E)) = \varnothing \cup \varnothing = \varnothing$.

Hence $L(D_a(E)) = D_a(L(E))$,

When $E = F + G$.

*Case (iv):*

Let $E = FG$ . Suppose if F and G are two regular expressions begin with a symbol 'a', then it can be found as in the case (iii), that

$$L(D_a(E)) = D_a(L(E)) = L(F')L(G').$$

Similarly, if F and G are regular expressions begin with other than 'a', it can be found as

$$L(D_a(E)) = D_a(L(E)) = \varnothing.$$

*Case (v):*

Let $E = F^*$, *then* $D_a(E) = D_a(F^*) = D_a(F)F^*$.

Again there are two possibilities, say

$F = aF'$ *or* $F = bF'$ *when* $a \neq b$.

In the first case, $D_a(F) = F'$ and $D_a(F) = F'F^*$.

In the second case, $D_a(F) = \varnothing$. Therefore $D_a(E) = \varnothing$.

Hence the statement $L(D_a(E)) = D_a(L(E))$ is trivially true.

If $D_a(F) = F'F^*$ , then

$$\begin{aligned} L(D_a(E)) &= L(F')L(F^*) \\ &= L(F')(L(F))^* \end{aligned}.$$

On the other hand, if $E = F^*$ and $F = aF' = aw$ . Then

$$L(E) = L(F^*) = L((aF')^*)$$

Hence

$$\begin{aligned} D_a(L(E)) &= D_a(L(aF')^*) \\ &= L(F')(L(F))^* \end{aligned}.$$

Hence $L(D_a(E)) = D_a(L(E))$.

This proves the theorem.

*3.11 Theorem*

Let $w = a_1 a_2 \ldots . a_n$, and E be a regular expression over an alphabet $\Sigma$. Then $D_w(E) = D_{a_n}(D_{a_1 a_2 a_3 \ldots . a_{n-1}}(E))$.

Suppose if $w = \varepsilon$ then $D_w E = E$ ,

As an illustration, let $w = aba$ and $E = (a+b)ab$ .then

$$\begin{aligned} D_w(E) &= D_{aba}((a+b)ab) \\ &= D_a(D_{ab}(a+b))ab \\ &= D_a(D_b(D_a(a) + D_a(b)))ab \\ &= D_a(D_b(\varepsilon + \varnothing))ab \\ &= D_a(\varnothing)ab = \varnothing ab = \varnothing \end{aligned}.$$

Generalizing the above illustration, the following theorems are obtained.

*3.12 Theorem*

Let $|w| = n$ and $\|E\| = m;\ n > m$ ; . then $D_w(E) = \varnothing$.

*3.13 Theorem*

If $E = w$ , then $D_w(E) = \varepsilon$ .

*3.14 Theorem*

If $w = au,\ u \in \Sigma^*$ , and $E = av,\ v \in \Sigma^*$ . Then $D_w(E) = D_u(v)$, *where* $a \in \Sigma$ .

Proof:

Let $u = a_1 a_2 \ldots . a_n$. Then

$$D_w(E) = D_{a_n}\left(D_{a_1a_2a_3....a_{n-1}}(E)\right)$$
$$= D_{a_n}\left(D_{a_{n-1}}\left(D_{a_1a_2a_3....a_{n-2}}(E)\right)\right)$$
$$= D_{a_n}\left(D_{a_{n-1}}\left(D_{a_{n-2}}....\right)\right)D_a(E)$$
$$= D_{a_n}\left(D_{a_{n-1}}\left(D_{a_{n-2}}....\right)\right)v$$
$$= D_{a_n}\left(D_{a_1....a_{n-1}}\right)v$$
$$= D_{a_1a_2a_3....a_n}(v)$$
$$= D_u(v)$$

*3.15 Theorem*

If $w = au;\ E = bv,\ where\ a,b \in \Sigma\ \ and\ u,v \in \Sigma^*$ ,
Then $D_w(E) = \varnothing$ .

## III. CONCLUSION

Brzozowski derivatives of the regular expressions are always helpful tool for constructing DFA. The generalizations of the derivatives are useful for transforming the size of the derivatives of the expressions research oriented work.